# Period Estimation in Astronomical Time Series Using Slotted Correntropy

Pablo Huijse, *Student Member, IEEE*, Pablo A. Estévez, *Senior Member, IEEE*, Pablo Zegers, *Senior Member, IEEE*, José C. Príncipe, *Fellow, IEEE*, and Pavlos Protopapas

*Abstract*—In this letter, we propose a method for period estimation in light curves from periodic variable stars using correntropy. Light curves are astronomical time series of stellar brightness over time, and are characterized as being noisy and unevenly sampled. We propose to use slotted time lags in order to estimate correntropy directly from irregularly sampled time series. A new information theoretic metric is proposed for discriminating among the peaks of the correntropy spectral density. The slotted correntropy method outperformed slotted correlation, string length, VarTools (Lomb-Scargle periodogram and Analysis of Variance), and SigSpec applications on a set of light curves drawn from the MACHO survey.

*Index Terms*—Correntropy, information theory, spectral analysis, time series analysis, variable stars.

## I. INTRODUCTION

RECENT advances in photometric technologies have facilitated the proliferation of extensive astronomical surveys such as MACHO [1], OGLE [2], and recently Pan-STARRS [3]. A light curve is a time series in which the measured phenomenon corresponds to the brightness (magnitude or flux) of a stellar object. Light curves are the basic tool for the analysis of variable stars [4], whose brightness varies through time due to internal physical processes, or to external factors such as interactions with other astronomical objects. Some variable stars, such as eclipsing binaries (EB), cepheids, and RR Lyrae, exhibit periodic behaviors that are reflected on their corresponding light curves. For example, EB stars are systems composed of two stars, whose brightness shows periodic variations due to the mutual eclipses between them. The period of a light curve is a key parameter for classifying variable stars [5], [6], and estimating other parameters such as mass and distance to Earth [7]. Light curves are unevenly sampled due to constraints on the observation schedules: the day-night cycle, weather conditions, calibration and repositioning of instruments. Several noise sources contaminate the measurements, e.g., the light of other stellar objects, atmospheric noise, instrument noise, etc.

The period of a time series can be estimated by analyzing the power spectral density (PSD) of its autocorrelation function. However, conventional autocorrelation estimators cannot be used directly if the time series is irregularly sampled. One option is to use the slotting technique [8], in which the time lags are defined as intervals. Some widely used techniques in astronomy for period estimation are the Lomb–Scargle (LS) periodogram [9], [10], Epoch Folding, Analysis of Variance (AoV) [11] and String Length (SL) methods [12]. The LS periodogram is an extension of the Fourier transform for unevenly sampled time series. In epoch folding, the time axis of the light curve is split into bins of size equal to a certain trial period. All bins are then plotted one on top of another. In analysis of variance the standard AoV statistic is computed over the binned and folded light curve. If the light curve is folded using its true period, the AoV reaches a minimum value. In SL methods, the light curve is folded using a trial period and the sum of distances between consecutive points in the folded curve is computed. The trial period with the shortest string length is taken as the underlying period.

In this work, we propose to combine the information theoretic concept of correntropy [13], [14] and signal processing techniques to design an automated method for period estimation in light-curves of periodic variable stars. In particular, we propose to estimate the fundamental period through the spectral analysis of the slotted correntropy. To discriminate the true period from its submultiples and spurious spectral peaks, a new information theoretic metric is proposed.

## II. BACKGROUND ON CORRENTROPY

Correntropy [14] is a generalization of the conventional correlation function. For a discrete-time strictly stationary univariate random process, $\{X_n, n \in 1, \ldots, N\}$, where $N$ is the length of the sequence, the autocorrentropy is defined as

$$V[m] = E[\kappa(X_n, X_{n-m})] \qquad (1)$$

where $E[\cdot]$ denotes the expectation operator, and $\kappa(\cdot, \cdot)$ is a kernel function that satisfies the Mercer's conditions [13]. In this letter, we consider the Gaussian kernel

$$G_\sigma(x_i - x_j) = \frac{1}{\sqrt{2\pi}\sigma} \exp\left(-\frac{\|x_i - x_j\|^2}{2\sigma^2}\right) \qquad (2)$$

where $x_i$ and $x_j$ are samples of the scalar random variables $X_i$ and $X_j$, respectively; and $\sigma$ is the kernel size.

As shown in [14], correntropy estimated with the Gaussian kernel (2) contains all even-order moments, and in particular the second-order moment that corresponds to the autocorrelation.

Manuscript received February 02, 2011; revised March 28, 2011; accepted March 30, 2011. Date of publication April 11, 2011; date of current version April 25, 2011. This work was supported by CONICYT-CHILE under Grants FONDECYT 1080643 and 1110701, and its Doctorate Scholarship program. The associate editor coordinating the review of this manuscript and approving it for publication was Dr. Alfred Mertins.

P. Huijse and P. A. Estévez are with the Electrical Engineering Department and Advanced Mining Technology Center, Universidad de Chile, Santiago, Chile (e-mail: phuijse@ing.uchile.cl, pestevez@cec.uchile.cl).

P. Zegers is with the College of Engineering and Applied Sciences, Universidad de los Andes, Santiago, Chile (e-mail: pzegers@miuandes.cl).

J. C. Príncipe is with the Computational Neuroengineering Laboratory, University of Florida, Gainesville, FL 32601 USA (e-mail: principe@cnel.ufl.edu).

P. Protopapas is with the IACS, School of Engineering and Applied Sciences and also with the Harvard-Smithsonian Center for Astrophysics, Harvard University, Cambridge, MA 02138 USA (e-mail: pprotopapas@cfa.harvard.edu).

Color versions of one or more of the figures in this paper are available online at http://ieeexplore.ieee.org.

Digital Object Identifier 10.1109/LSP.2011.2141987

The kernel size controls the emphasis given to higher-order moments over the second moment. The sample mean can be used to estimate the correntropy as

$$\hat{V}[m] = \frac{1}{N-m+1} \sum_{n=m}^{N-1} G_\sigma(x_n - x_{n-m}) \quad (3)$$

for $0 < m < N-1$. The name correntropy comes from the fact that it looks like correlation, but the sum of (3) over the lags is the information potential (IP) in information theoretic learning [13], which is defined as

$$IP(\{x_n\}) = \frac{1}{N^2} \sum_{i=0}^{N-1} \sum_{j=0}^{N-1} G_\sigma(x_i - x_j). \quad (4)$$

The negative logarithm of the IP is the Renyi's quadratic entropy estimated through Parzen windows [13]. For large values of the kernel size the IP tends towards a scaled and biased version of the variance.

The correntropy spectral density (CSD) is defined as

$$P[f] = \sum_{m=-\infty}^{\infty} \left(\hat{V}[m] - \langle \hat{V}[m] \rangle \right) e^{-j2\pi(f/F_s)m} \quad (5)$$

where $\langle \hat{V}[m] \rangle$ is the sample mean estimator of correntropy over the lags and $F_s$ is the sampling frequency. The CSD is the Fourier transform of the centered autocorrentropy and it can be considered as a generalized PSD function. For further details on correntropy and its interpretation see [13].

### III. SLOTTED CORRENTROPY AND IP-BASED CRITERION

In this letter, we propose estimating correntropy directly from an irregularly sampled time series using slotted time lags. Slotted correntropy is computed as

$$\hat{V}[k\Delta\tau] = \frac{\sum\limits_{i,j}^{N} G_\sigma(x_i - x_j) B_{k\Delta\tau}(t_i, t_j)}{\sum\limits_{i,j}^{N} B_{k\Delta\tau}(t_i, t_j)} \quad (6)$$

for $k = 0, 1, 2, \ldots, [\tau_{\max}/\Delta\tau]$, where $[\cdot]$ is the nearest integer function, $\Delta\tau$ is the slot size, $\tau_{\max}$ is the maximum lag, $t_i$ and $t_j$ are the times of samples $x_i$ and $x_j$, respectively; and

$$B_{k\Delta\tau}(t_i, t_j) = \begin{cases} 1, & \text{if } |(t_i - t_j) - k\Delta\tau| < 0.5\Delta\tau \\ 0, & \text{otherwise.} \end{cases} \quad (7)$$

The slotted correntropy is sampled at intervals of length $\Delta\tau$.

In addition, we propose a new period discrimination metric, in order to extract the peak in the CSD associated with the fundamental period among spurious and multiple/submultiple peaks. A trial period is considered spurious if it is neither the true period nor any of its integer multiples and submultiples. The procedure to compute the IP-based discrimination metric for a certain trial period $P_c$ is as follows.

1) Transform the time axis of the light-curve as follows

$$\phi_n(P_c) = \frac{(t_n \bmod P_c)}{P_c}, \quad n = 1, \ldots, N$$

where mod stands for the modulo operation. With this procedure a folded light curve $\{\phi_n(P_c), x_n\}$ is obtained.

2) Smooth the folded light curve by taking a moving average of 20 samples. Search for local maxima and minima using samples in windows of size $M = N/10$ and an overlap of half the window size. The value of $M$ is set as a compromise between detecting spurious peaks as local maxima and missing true peaks.

3) Segment the folded light curve (non-smoothed) into bins so that each local optima corresponds to the center of a different bin. The boundaries of the bins are chosen as the mid-points between adjacent local optima. This procedure results in an adaptive number of bins $H$.

4) Compute the discrimination metric as the average of the squared differences between the IP of each individual bin, $h$, and the global IP:

$$Q(P_c) = \frac{1}{H} \sum_{h=1}^{H} \left[IP(\{x_n\}_{n \in h}) - IP(\{x_n\})\right]^2. \quad (8)$$

5) Repeat steps 1–4 for each trial period. The best period is selected as the one that maximizes the IP metric (8).

Steps 2 and 3 are referred later as dynamic binning. When maximizing (8) we are searching for the largest difference between the information content in the dynamically chosen bins and that of the complete light curve. The maximum occurs when the light curve is folded using its underlying period.

### IV. METHODS

A subset of 600 light-curves was drawn from the MACHO survey [1]. It contains 200 light-curves from each of the following three types of variable stars: EBs, cepheids and RR Lyrae, whose periods range from 0.2 days to 200 days. Each light-curve has approximately 1000 samples and contains three data columns: time, magnitude and an error estimation for the magnitude. The periods of these light-curves were estimated by expert astronomers from the Harvard Time Series Center (TSC) using epoch folding, AoV, and visual inspection. In this letter, we consider the TSC periods to be the gold standard.

We propose the following method for automatic period estimation in light curves.

1) Normalize the light curve's magnitude to have zero mean and unit standard deviation. Discard samples having an error estimation greater than the mean error plus two times its standard deviation (usually less than 1% of the samples of a light curve are discarded).

2) Select a window of half the length of the light curve containing the maximum number of samples per day. The selected window and the whole light curve are independently analyzed in the next stages in order to generate the pool of trial periods.

3) Compute the slotted correntropy using (6). The maximum lag is set to 0.1 N, where N is the light curve length. This value is chosen as a trade-off between having enough samples to estimate correntropy with longer lags and bounding the longer period that could be detected.

4) Compute centered slotted correntropy by removing its mean value in the interval $[-0.1N, 0.1N]$. Multiply this result by a Hamming window in the same interval, and compute the CSD.

5) Store the periods associated with the $N_p$ highest peaks of the CSD. For each trial period $P_c$ compute the IP metric in the interval $[P_c - 0.5, P_c + 0.5]$ with step size 1e-3, by applying steps 1–5 described in Section III. Save the fine-tuned period that maximizes (8).
6) Save the fine-tuned period with highest $Q$ among all the periods obtained from both windows selected in Step 2.

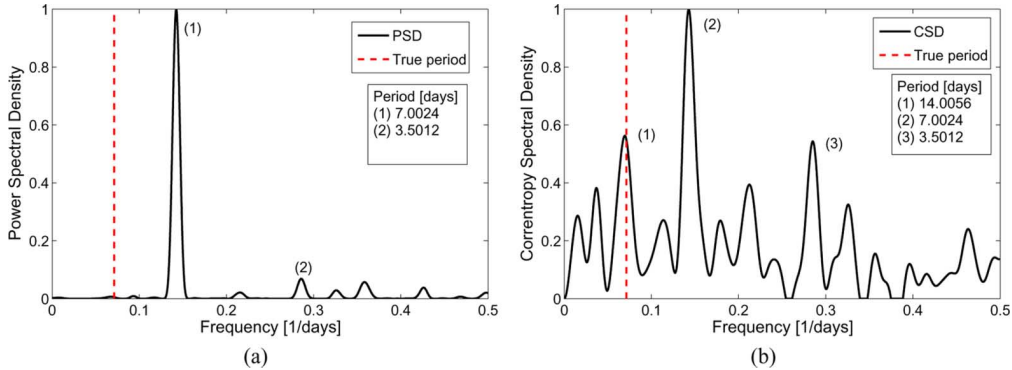

Fig. 1. Light-curve analysis of object 1.3449.948 from the MACHO survey. The dashed lines mark the position of the known period. a) PSD of the slotted autocorrelation function. The true period appears very attenuated. b) CSD of the slotted autocorrentropy function for a kernel size of 0.1. The true period appears as the second largest peak.

Our method has the following user-defined parameters:
- **Gaussian kernel size** $\sigma$: If $\sigma$ is set too large the higher-order moments in correntropy will be ignored. If it is set too small then correntropy will not be able to discriminate between signal and noise. Steps 1–6 of our method are repeated for 25 values of the kernel size in the interval $[0.01, 5]$. Each kernel size provides $N_p$ trial periods.
- **Correntropy slot size** $\Delta\tau$: The slot size defines the time lag resolution of correntropy. If it is set too small, it will be harder to satisfy condition (7). If it is set too large short periods may not be found. We set $\Delta\tau = 0.25$ days to capture the shorter periods present in the data set.
- **Number of peaks analyzed from the CSD** $N_p$: In our experiments we found that setting $N_p = 10$ is a good trade-off between obtaining the biggest hit rate and having less computational load.

## V. EXPERIMENTAL RESULTS

### A. Analysis of a Single Light-Curve

We illustrate our method by using as an example the light curve 1.3449.948 from the MACHO survey, which has a reported period of 14.0063 days. As a reference the PSD of slotted correlation is illustrated in Fig. 1(a), with a global maximum at 7.0024 days, half of the true period. The true period appears to be very attenuated in the PSD [dashed vertical line in Fig. 1(a)], even smaller than spurious peaks. The CSD of slotted correntropy is illustrated in Fig. 1(b). The true period appears clearly as the second highest peak in the CSD. The IP metric allows us to correctly select the second highest peak as the fundamental period, because $Q(14.0055) = 0.1288$ is larger than $Q(7.0024) = 0.0984$.

Fig. 2 shows a contour plot of the CSD as a function of the frequency and kernel size. This plot shows that for a kernel size greater than $\sigma = 1$, only the peak associated with half the true period remains in the spectrum. This result supports the theory that when $\sigma$ is large, correntropy approximates correlation and the CSD approximates the PSD [13]. On the other hand, for a kernel size smaller than $\sigma = 0.05$, several submultiples but also some spurious peaks are more outstanding than the true period (Fig. 2 cut c).

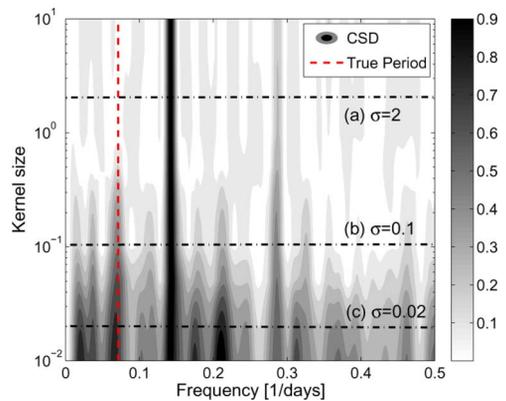

Fig. 2. CSD of light curve 1.3449.948 as a function of frequency and kernel size. In cut (a) correntropy and CSD approximate correlation and PSD, respectively (see Fig. 1(a)). In cut (b) the true period appears as the second highest peak in the spectrum (see Fig. 1(b)). In cut (c) the true period is the fourth highest peak, even lower than a spurious peak.

### B. Comparison Between AoV, SL and IP-Based Metric

Estimation results are classified as hits, multiples and misses by comparing them with the true periods reported by the TSC team. A given period is considered a hit if its relative difference from the true period is less than 0.5%; a multiple if its difference with any integer multiple/submultiple of the true period is less than 0.5%; otherwise it is a miss. Two versions of the IP-based metric are compared: fixed binning and dynamic binning. The IP-based metric is compared to AoV and the string length method using the Lafler–Kinman statistic (SLLK) [12]. In our first experiment all three metrics are applied to discriminate the correct period among the top 10 peaks of the CSD using 200 EB light-curves. Table I shows the results obtained by using the different strategies. The dynamic binning strategy allows obtaining more hits than the fixed binning strategy, achieving an increase of 3.5% to 6% in the hit rate. The IP-based metric obtained 3.5% more hits than the AoV criterion and 14% more hits than SLLK.

### C. Comparison With Related Methods

The performance of the proposed method was compared with widely used applications for period estimation in astronomy. We considered two software solutions available on the internet: VarTools [15] and SigSpec [16]. VarTools includes LS and AoV light curve analyses. SigSpec combines Fourier based methods with statistical metrics of spectral significance. We also considered the SLLK string length method [12]. For Vartools LS, the highest peak of the periodogram gives the estimated period. In Vartools LS + IP, the IP metric is used to discriminate among the top 10 peaks obtained with the LS periodogram. Likewise, in SLLK + IP, the IP metric is used to discriminate among the top 10 smallest strings found by SLLK. For SLLK and Vartools AoV, the corresponding statistics are minimized in an array of periods ranging from 0.2 to 200 days with a step size of 1e-4. For slotted $\text{correlation} + \text{IP}$ we compute the top 10 peaks of the PSD, and then apply the IP metric to discriminate among them.

TABLE I
COMPARISON BETWEEN DISCRIMINATION METRICS APPLIED TO THE TOP TEN PEAKS OF CSD, USING A SUBSET OF 200 EB LIGHT CURVES FROM THE MACHO SURVEY

| Discrimination metric | Hits[%] | Multiples[%] | Misses[%] |
|---|---|---|---|
| AoV/fixed binning (10 bins) | 67.0 | 32.5 | 0.5 |
| AoV/dynamic binning | 70.5 | 29.0 | 0.5 |
| IP metric/fixed binning (10 bins) | 68.0 | 31.5 | 0.5 |
| IP metric/dynamic binning | 74.0 | 25.5 | 0.5 |
| SLLK | 60.0 | 37.5 | 2.5 |

TABLE II
PERFORMANCE OF THE PROPOSED PERIOD ESTIMATOR (Slotted Correntropy + IP) VERSUS CONVENTIONAL TECHNIQUES IN A SUBSET OF 200 EB LIGHT CURVES FROM THE MACHO SURVEY

| Period estimation methods | Hits[%] | Multiples[%] | Misses[%] |
|---|---|---|---|
| Slotted correntropy + IP | 74.0 | 25.5 | 0.5 |
| Slotted correlation + IP | 50.0 | 48.5 | 1.5 |
| VarTools LS | 11.0 | 89.0 | 0.0 |
| VarTools LS + IP | 18.0 | 82.0 | 0.0 |
| VarTools AoV | 39.5 | 60.5 | 0.0 |
| SigSpec | 11.0 | 88.5 | 0.5 |
| SLLK | 42.5 | 54.5 | 3.0 |
| SLLK +IP | 65.0 | 34.5 | 0.5 |

TABLE III
PERFORMANCE OF THE PROPOSED PERIOD ESTIMATOR (Slotted Correntropy + IP) VERSUS CONVENTIONAL TECHNIQUES IN A SUBSET OF 400 CEPHEIDS AND RRL LIGHT CURVES FROM THE MACHO SURVEY

| Period estimation methods | Hits[%] | Multiples[%] | Misses[%] |
|---|---|---|---|
| Slotted correntropy + IP | 97.00 | 2.75 | 0.25 |
| Slotted correlation + IP | 93.00 | 5.75 | 1.25 |
| VarTools LS | 97.00 | 2.75 | 0.25 |
| VarTools LS + IP | 97.00 | 2.75 | 0.25 |
| VarTools AoV | 97.00 | 2.75 | 0.25 |
| SigSpec | 95.50 | 4.25 | 0.25 |
| SLLK | 68.50 | 28.00 | 3.50 |
| SLLK +IP | 90.25 | 4.25 | 0.25 |

Table II shows the results obtained with each of the described methods in a subset of 200 EB light curves. The slotted $\text{correntropy} + \text{IP}$ method obtained the highest hit rate (74%), followed by $\text{SLLK} + \text{IP}$ (65%), and slotted $\text{correlation} + \text{IP}$ (50%). Table III shows the results obtained in a subset of 400 cepheids and RR Lyrae light curves. The slotted $\text{correntropy} + \text{IP}$ method, Vartools LS and Vartools AoV obtained the highest hit rate (97%), followed by SigSpec (95.5%). SLLK enhanced by the IP metric achieved 90.25%.

## VI. CONCLUSIONS

An automated method for period estimation in light curves of periodic variable stars has been proposed and tested. The method is based on the spectral analysis of the slotted correntropy. The proposed method obtained a hit rate of 74% in a subset of 200 light curves from eclipsing binaries from the MACHO survey, outperforming slotted correlation (50.0%) and conventional methods. Our method obtained a hit rate of 97% in a subset of 400 light curves from cepheids and RR Lyrae stars, where Vartools LS and AoV achieved the same performance. A new information theoretic metric for trial period discrimination has been proposed, which is computed on dynamically binned folded light curves. The IP criterion performed better than AoV and SLLK in the trial period discrimination task. Future work will include among other issues, devising a better metric for trial period discrimination, and testing our method on massive astronomical databases.

## REFERENCES

[1] C. Alcock et al., "The MACHO project: Microlensing results from 5.7 years of LMC observations," *Astrophys. J.*, vol. 542, pp. 281–307, 2000.
[2] A. Udalski, M. Kubiak, and M. Szymanski, "Optical gravitational lensing experiment. OGLE-2—The second phase of the OGLE project," *Acta Astron.*, vol. 47, pp. 319–344, 1997.
[3] N. Kaiser et al., "Pan-STARRS: A large synoptic survey telescope array," in *Soc. Photo-Optical Instrum. Eng. (SPIE) Conf. Ser.*, 2002, vol. 4836, pp. 154–164.
[4] M. Petit, *Variable Stars*. New York: Wiley, 1987.
[5] J. Debosscher, L. M. Sarro, C. Aerts, J. Cuypers, B. Vandenbussche, R. Garrido, and E. Solano, "Automated supervised classification of variable stars. I. Methodology," *Astron. Astrophys.*, vol. 475, pp. 1159–1183, 2007.
[6] G. Wachman, R. Khardon, P. Protopapas, and C. Alcock, "Kernels for periodic time series arising in astronomy," in *Proc. Eur. Conf. Machine Learning*, 2009, vol. 5782, pp. 489–505.
[7] D. M. Popper, "Stellar masses," *Annu. Rev. Astron. Astrophys.*, vol. 18, pp. 115–164, 1968.
[8] W. T. Mayo, "A discussion of limitations and extentions of power spectrum estimation with burst-counter LDV systems," in *Proc. Second Int. Workshop on Laser Velocimetry*, 1974, pp. 90–104.
[9] N. R. Lomb, "Least-squares frequency analysis of unequally spaced data," *Astrophys. Space Sci.*, vol. 39, pp. 447–462, 1976.
[10] J. D. Scargle, "Studies in astronomical time series analysis. II. Statistical aspects of spectral analysis of unevenly spaced data," *Astrophys. J.*, vol. 263, pp. 835–853, 1982.
[11] A. Schwarzenberg-Czerny, "On the advantage of using analysis of variance for period search," *Monthly Notices of the Royal Astron. Soc. (MNRAS)*, vol. 241, pp. 153–165, 1989.
[12] D. Clarke, "String/rope length methods using the Lafler-Kinman statistic," *Astron. Astrophys.*, vol. 386, pp. 763–774, 2002.
[13] J. C. Principe, *Information Theoretic Learning: Renyi's Entropy and Kernel Perspectives*. New York: Springer Verlag, 2010.
[14] I. Santamaría, P. P. Pokharel, and J. C. Principe, "Generalized correlation function: Definition, properties, and application to blind equalization," *IEEE Trans. Signal Process.*, vol. 54, pp. 2187–2197, 2006.
[15] J. D. Hartman et al., "Deep MMT transit survey of the open cluster M37. II. Variable stars," *The Astrophysical Journal*, vol. 675, pp. 1254–1277, 2008.
[16] P. Reegen, "SigSpec I. frequency- and phase-resolved significance in Fourier space," *Astronomy & Astrophysics*, vol. 467, pp. 1353–1371, 2007.